\documentclass[apj]{emulateapj}

\usepackage{enumerate}
\usepackage{graphicx}
\usepackage{txfonts}
\usepackage{natbib}
\usepackage{verbatim}
\usepackage[colorlinks,urlcolor=blue,citecolor=blue,linkcolor=blue]{hyperref}

\shorttitle{Measurements of EUV Coronal Holes and Open Magnetic Flux}
\shortauthors{Lowder et al.}

\begin{document}

\title{Measurements of EUV Coronal Holes and Open Magnetic Flux}
\author{C. Lowder, J. Qiu, R. Leamon}
\affil{Department of Physics, Montana State University, Bozeman, Montana 59717}
\email{clowder@solar.physics.montana.edu}
\and
\author{Y. Liu}
\affil{W. W. Hansen Experimental Physics Laboratory, Stanford University, Stanford, CA 94305}

\begin{abstract}

Coronal holes are regions on the Sun's surface that map the foot-prints of open magnetic field lines. We have developed an automated routine to detect and track boundaries of long-lived coronal holes using full-disk EUV images obtained by SoHO:EIT, SDO:AIA, and STEREO:EUVI. We measure coronal hole areas and magnetic flux in these holes, and compare the measurements with calculations by the PFSS model. It is shown that, from 1996 through 2010, the total area of coronal holes measured with EIT images varies between 5\% and 17\% of the total solar surface area, and the total unsigned open flux varies between 2-5$\times 10^{22}$ Mx. The solar cycle dependence of these measurements are similar to the PFSS results, but the model yields larger hole areas and greater open flux than observed by EIT. The AIA/EUVI measurements from 2010-2013 show coronal hole area coverage of 5-10\% of the total surface area, with significant contribution from low latitudes, which is under-represented by EIT. AIA/EUVI have measured much enhanced open magnetic flux in the range of 2-4$\times 10^{22}$ Mx, which is about twice the flux measured by EIT, and matches with the PFSS calculated open flux, with discrepancies in the location and strength of coronal holes. A detailed comparison between the three measurements (by EIT, AIA-EUVI, and PFSS) indicates that coronal holes in low latitudes contribute significantly to the total open magnetic flux. These low-latitude coronal holes are not well measured with either the He I 10830 line in previous studies, or EIT EUV images; neither are they well captured by the static PFSS model. The enhanced observations from AIA/EUVI allow a more accurate measure of these low latitude coronal holes, and their contribution to open magnetic flux.

\end{abstract}


\section{Introduction}

Coronal holes are observationally a region of diminished emission in the extreme ultraviolet (EUV) and X-ray wavelengths, as compared to the quiet sun background level. The earliest observations of coronal holes were made in EUV wavelengths. \cite{1968IAUS...35..411T} noted from spectroheliograms obtained by rocket experiments that EUV emission in polar regions seemed weaker than in surrounding regions. More detailed spectroscopic observations were made possible by instruments onboard Skylab, which yielded some of the early measurements of the chromosphere, transition region, and corona properties in coronal holes \citep{1974ApJ...194L.115H}. Coronal holes have since then been observed in many wavelengths from radio, near infrared (He I 10830 line), white-light, to EUV and X-rays both on disk and from the limb, and their properties have been extensively studied, including the temperature, density, flow velocity, energy flux, lifetime, and magnetic fields. Detailed reviews of the plasma and magnetic properties of coronal holes have been given by \citet{1977RvGSP..15..257Z, 1979SSRv...23..139H, 2009LRSP....6....3C}.

There has been long-standing interest in studying coronal holes because of their association with large scale solar magnetic fields and solar winds. Coronal holes are considered to map regions on the Sun's surface where magnetic field lines are open to the heliosphere \citep[][and references therein]{2009SSRv..144..383W}. Simply put, as coronal plasmas and energy flow outward along open magnetic field lines, the coronal density in these regions is decreased, resulting in diminished X-ray and EUV emission. Long-lived coronal holes that exist for days or months are often found in high-latitude regions, such as polar caps. Some of the long-lived coronal holes also extend into low-latitude regions. They are usually associated with fast solar wind in a quasi-steady state, as originally proposed by \citet{1958ApJ...128..664P}. In addition to long-lived holes, there are regions of depleted EUV emission that evolve rapidly, expanding and refilling in a matter of hours. These transient coronal holes, or coronal dimmings, are most likely associated with a depletion of coronal material following a coronal mass ejection (CME) which is magnetically anchored in these regions \citep{1983SSRv...34...21R, 2000GeoRL..27.1431T, 2008ChJAA...8..329Y, 2009ApJ...706..376A}. 

Models have been developed to study the large scale magnetic fields on the Sun and to calculate the open magnetic flux budget in the heliosphere. These models include extrapolation models, such as the widely used Potential Field Source Surface (PFSS) model \citep{1969SoPh....6..442S, 1992ApJ...392..310W}, as well as more sophisticated MHD models \citep[See review by][]{2012LRSP....9....6M}. All models use observed photospheric magnetograms or some variation of them \citep{2003SoPh..212..165S} as the boundary condition. These models compute open magnetic field lines that extend to the heliosphere, and calculate the total open flux to compare with in-situ measurements of Interplanetary Magnetic Field (IMF) by satellites such as Ulysses \citep{1992A&AS...92..221B, 2012LRSP....9....6M}. 

To validate models, the foot-prints of model computed open field lines on the Sun's surface are sometimes compared with observed coronal holes. It was shown that the global pattern of the open-field foot-points computed by either MHD or PFSS models during one solar rotation in general matches the long-lived coronal hole boundaries observed in the chromosphere He I 10830 line or EUV/X-rays \citep{1982SoPh...79..203L, 1996Sci...271..464W, 1998JGR...10314587N}, in particular in polar caps. However, to our knowledge, systematic measurement of magnetic flux directly from observed coronal holes has been done rarely; \citet{2002SoPh..211...31H} provides one of the very few direct measurements of open flux from polar holes detected in the He I 10830 line covering a decade from 1990 to 2000. Furthermore, more than thirty years ago, \citet{1979SSRv...23..139H} raised the question of the relative contribution of magnetic open flux by coronal holes of all kinds, including strong polar holes, weak low-latitude holes, and possibly miniature coronal holes in young active regions. Recently, it was shown that low-latitude long-lived holes contribute significantly to the open flux during the solar maximum \citep{2002JGRA..107.1488N, 2002JGRA..107.1154L, 2003SoPh..212..165S}, and it was also proposed that contribution by some short-lived coronal holes associated with CMEs may not be negligible \citep{2007ApJ...667L..97R}. Note that these studies applied PFSS or MHD models, as well as inference from the in-situ IMF measurements, to reach these conclusions. It is important to test these suggestions from observations of coronal holes and direct measurements of magnetic flux in the holes at different latitudes.

The \textit{Extreme-ultraviolet Imaging Telescope} \citep[EIT;][]{1995SoPh..162..291D} onboard the \textit{Solar and Heliospheric Observatory} (SoHO) has been observing the corona since 1995, and therefore provides a stable database suitable for systematic tracking and characterization of EUV coronal holes for the past one and and one-half solar cycle. In 2010, the \textit{Solar Dynamics Observatory} (SDO) was launched; meanwhile, the \textit{Solar Terrestrial Relations Observatory} (STEREO) A and B have reached vantage points that are separated from SDO by nearly $\pm 90$ degrees. Therefore, the EUV telescope \textit{Atmospheric Imaging Assembly} \citep[AIA;][]{2012SoPh..275...17L} onboard SDO in conjunction with the \textit{Extreme Ultraviolet Imager} \citep[EUVI;][]{2008SSRv..136...67H} onboard STEREO A and B have been able to provide nearly full coverage of the solar surface observed in EUV wavelengths. These data can be employed to obtain continuous, consistent, full solar surface observations of coronal hole boundaries for the present solar cycle 24.

In this paper, we utilize these available databases to track and characterize EUV coronal holes, and make a comparison of magnetic flux directly measured from these holes with open flux computed by the widely used PFSS model. The analysis and model are performed for the past solar cycle (1996 - 2011) observed by EIT, as well as the two years when the Sun has been observed by EIT, AIA, and EUVI. Because a large amount of data over a long period are analyzed, we limit the scope of the present study to a cadence of 12 hours for AIA/EUVI observations and 24 hours for EIT observations; therefore, the focus is on relatively long-lived coronal holes, but not transient holes. In the following text, we will describe the data and analysis method in Section 2. Section 3 will present the results of coronal hole tracking and measurements of magnetic flux from the holes in comparison with the PFSS model. Discussions and conclusions are given in Section 4.

\section{Methodology}
To analyze coronal holes, we employ the EUV full-disk images obtained by EIT from 1996 May to the end of 2010 and by AIA and EUVI since 2010 May. Photospheric magnetic field measurements are obtained by Michelson Doppler Imager \citep{1995SoPh..162..129S} onboard SoHO from 1996 to 2010, and then by Helioseismic and Magnetic Imager \citep{2012SoPh..275..207S} since 2010 April. We develop an automated procedure to detect coronal holes using these data, and study some properties of coronal holes, including their persistency, latitude distribution, and magnetic flux. These properties are then compared with the PFSS model results.

\subsection{Observations}

\begin{deluxetable*}{rrrrrr}
\tablecolumns{6}
\tablewidth{0pt}
\tablecaption{Data coverage for each instrument source}
\tablehead{ 
\colhead{}    &  \colhead{} & \multicolumn{2}{c}{Start}  & \multicolumn{2}{c}{End} \\ 
\cline{3-4} \cline{5-6} \\ 
\colhead{Source} & \colhead{Observable} & \colhead{Carrington Rotation} & \colhead{Date} & \colhead{Carrington Rotation} & \colhead{Date}}
\startdata 
SoHO:EIT					 & EUV 195\AA & 1909.96 & 1996 05 31 & 2105.27 & 2010 12 31\\
SDO:AIA	 				& EUV 193\AA & 2096.76 & 2010 05 13 & 2133.17 & 2013 01 30\\
STEREO:EUVI			 	& EUV 195\AA & 2096.76 & 2010 05 13 & 2133.17 & 2013 01 30\\
WSO Harmonic Coefficients	& Radial magnetic field & 1893.00 & 1995 02 23 & 2113.00 & 2011 07 29\\
SoHO:MDI Synoptic Charts	& Radial magnetic field& 1911.00 & 1996 06 28 & 2104.00 & 2010 11 26\\
SDO:HMI Synoptic Charts		& Radial magnetic field & 2096.00 & 2010 04 22 & 2134.00 & 2013 02 21
\enddata 
\label{table:table1}
\end{deluxetable*}

\begin{figure}[h!tbp]
\centering
\includegraphics[width=7.5cm]{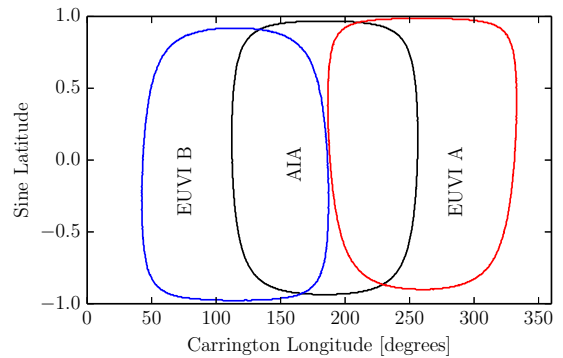}
\caption{Available instrument coverage for 2010 June 29. Contours are marked for each instrument field of view, truncated at 0.95 R$_\odot$ for purposes of illustration. The fields of view for AIA, EUVI:A, and EUVI:B are centered at Carrington longitude values of 190, 265, and 120 degrees, respectively. Note the enhanced polar viewing angles achieved with the EUVI instruments.}
\label{fig:f1}
\end{figure}

Table~\ref{table:table1} displays the availability of each dataset in use. As mentioned previously, one of the benefits of using a combination of data sources from AIA and EUVI A/B is the ability to have near full surface observations. For the example observation provided, 2010 June 29, the two STEREO spacecraft have separated in their orbits to provide a nearly full surface view. Several months later, full coverage was achieved by the three spacecraft, and this complete coverage will continue for several years to come. Figure~\ref{fig:f1} displays the coverage overlap between the three instruments. Contours are marked at 0.95 R$_\odot$ for the field of view of each instrument. AIA, EUVI:A, and EUVI:B are centered at Carrington longitude 190, 265, and 120 degrees, respectively.

The brief overlap between the EIT and AIA-EUVI datasets will serve as a crucial comparison between the two sources. EUVI and EIT observations in 195$\AA$~measure emission primarily from Fe XII. AIA observations in 193$\AA$~measure primarily emission from Fe XII, but with additional emission from Fe XXIV whose formation temperature is about 10 MK. Since coronal holes are characterized by low-temperature plasmas \citep{2009LRSP....6....3C}, the difference in the wavelengths between EIT and AIA-EUVI is unlikely to bias the coronal hole detection. On the other hand, detection of the boundaries of coronal holes, especially relatively smaller holes, may be affected by scattered light from adjacent bright features. EIT images may be subject to a higher level of scattered light \citep{2012ApJ...749L...8S}. The effect of the scattered light in different instruments will manifest itself when we make the comparison of hole detection, and will be discussed later.

Accompanying the EUV observations by the above-mentioned instruments, MDI and HMI provide the magnetic field measurements. Processed charts of radial magnetic field strength are employed from both MDI and HMI observations. MDI charts are corrected to account for missing polar field information due to inconvenient solar tilt angles throughout the year. The details of this correction are discussed in \citet{2011SoPh..270....9S}. Considering the pass-off of data from MDI to HMI in our studies of open magnetic flux, validating observations between these two instruments is crucial. Effects such as instrument degradation and sensitivity are key. A more detailed study of the inter-calibration between the two measurements is available from the work of \citet{2012SoPh..279..295L}.

\subsection{Thresholding routine}

Building on the work of \citet{2009SoPh..256...87K}, we've developed an enhanced routine for automated coronal hole detection, capable of working with multiple input sources of data. The routine first identifies dark features in EUV images, and then utilizing synoptic maps of calculated radial magnetic field, this routine is able to distinguish between coronal holes (dominated by one magnetic polarity) and filament channels (mixed polarity).

\begin{figure}[h!tbp]
\includegraphics[width=8cm]{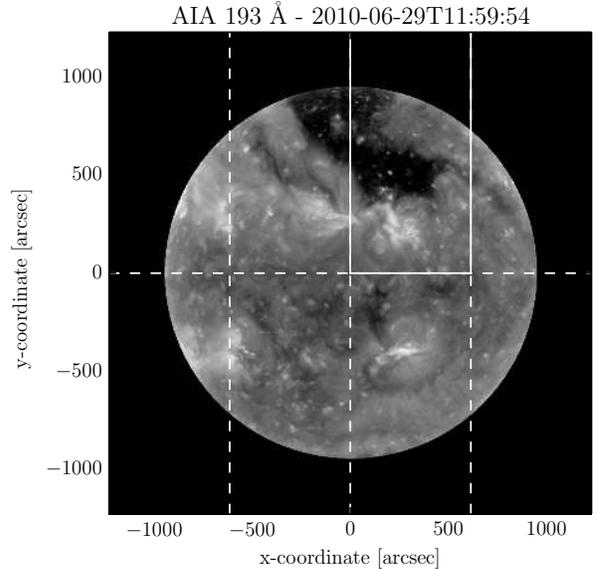}
\caption{Full-disk EUV data taken at 193$\AA$~by the AIA instrument aboard SDO. The thresholding routine proceeds to partition this data into eight sub-arrays, on each of which the code runs the thresholding calculation. Each sub-array contains a differing mixture of bright and dark features. The solid boxed sub-array marks the sub-array being considered in Figure~\ref{fig:f3}.}
\label{fig:f2}
\end{figure}

Each EUV image is processed using standard SSWIDL software routines for each particular instrument. The AIA images are read into memory and processed using \verb+read_sdo.pro+ and \verb+aia_prep.pro+, respectively. The EUVI:A/B images are read and processed using the \verb+secchi_prep.pro+ routine. The EIT images are read and processed using the \verb+eit_prep.pro+ routine. A mask is applied to each respective image to remove off-limb information. Each of the processed and cropped EUV images is subdivided into eight sub-arrays for further analysis as described below. Figure~\ref{fig:f2} illustrates this sub-array arrangement. 

\begin{figure}[h!tbp]
\centering
\includegraphics[width=8cm]{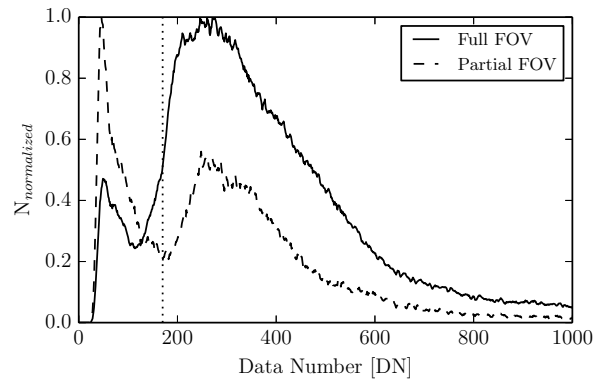}
\caption{Histogram of the sub-array marked in Figure~\ref{fig:f2}, calculated in terms of data number for the recorded image. This figure displays a limited range of the full extent of the data, and has been boxcar smoothed with a width of 10 DN for clarity of the underlying form of the data. The solid curve and dashed curve refer to the full field of view, and the marked subarray, respectively. The vertical line indicates the local minimum value that is appropriate as a thresholding value in DN. Note that this local minimum is not always readily detectable in the full field of view data.}
\label{fig:f3}
\end{figure}

\citet{1998A&A...335..733G} have shown that a histogram of EUV intensity corresponds to the contribution from multiple sources. More importantly, by thresholding an image in the valley between contributing peaks in EUV intensity histograms, features can be separated. Following this experience, we use the EUV intensity histogram to define the threshold for coronal holes. It was also suggested that this method works better by partitioning a full-disk EUV image into a few sub-frames \citep{2009SoPh..256...87K}. Partitioning reduces the overlap between features in an intensity histogram, as there are simply fewer contributing sources, particularly the quiet sun emission, in the narrow field of view. Consider a histogram of EUV intensity, measured in DN, as displayed in Figure~\ref{fig:f3}. The EUV intensity histograms of the full field of view and the sub-array are displayed in the solid curve and dashed curve, respectively. Each curve is normalized for comparison.

The histogram of the partial frame exhibits sharper peaks than in the full histogram. The two peaks seen in the partial field of view at 50 DN and 250 DN correspond to coronal holes and quiet sun, respectively. The local minimum between the quiet sun peak and coronal hole peak defines the threshold between these two features. By plotting the contour at this threshold on the original EUV image, we find this threshold rather reliably outlines the coronal hole boundary. In contrast, the full field of view histogram displays a very broad and slightly shifted peak, pushing against the lower-intensity coronal hole peak, indicative of an over-dominant contribution by the quiet sun emission. By partitioning the image into sub-arrays, particularly in those sub-frames where coronal holes are present, the quiet-sun contribution is minimized. In addition, the shape of each sub-array is vertical in nature to better capture contribution from both polar coronal holes and quiet regions.

An EUV intensity histogram is calculated for each sub-array, and numerical first and second derivatives are taken. The quiet sun DN value is defined from the median of the on-disk image. From our experiments, the coronal hole threshold is close to half of the quiet sun DN value. The routine searches for a local minimum closest to half of the quiet sun DN value in each sub-array to define the threshold value. If a local minimum does not exist within 0.3-0.7 DN$_{QS}$, the sub-array is discarded. After each threshold value is computed a full field of view thresholding value is computed from the mean of the valid sub-array threshold values. This final threshold is then applied to the full FOV image to find coronal hole boundaries.

\begin{figure}[h!tbp]
\centering
\includegraphics[width=7.5cm]{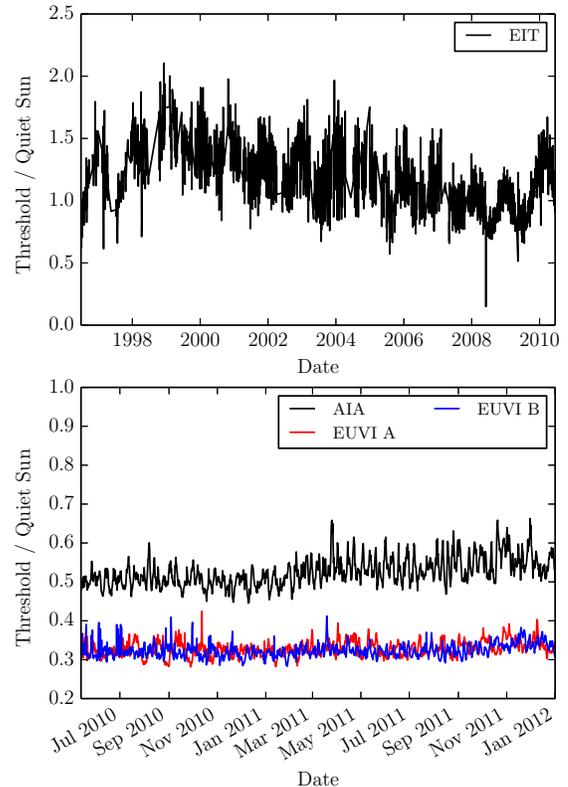}
\caption{Ratio of the data number thresholding value to the quiet sun value for each EIT in the upper panel, and AIA, EUVI A/B in the lower panel.}
\label{fig:f4}
\end{figure}

This code was then tested with and used to gather results from several instruments, AIA, EUVI-A and B, and EIT. Despite using multiple instruments with differing calibrations, this modified routine has proven consistent with maintaining relatively stable coronal hole thresholding values as well as coronal hole boundaries over a long time period. Consider Figure~\ref{fig:f4}, which displays the thresholding value for EIT, AIA, EUVI-A, and EUVI-B. The values are displayed as a ratio to the quiet sun value, to avoid dimensionality. Quiet sun values were determined through the median data number within each frame. This ratio value stays relatively stable throughout the dataset shown.

It is seen that the threshold relative to the quiet sun intensity is markedly different for different instruments. The mean threshold to quiet sun ratio is 0.53, 0.33, 0.32, and 1.18, for AIA, EUVI A, EUVI B, and EIT, respectively. The difference must be partly due to the different instrument calibration, sensitivity, and scattered light level. It is noted that the threshold for EIT is rather large, perhaps due to higher scattered light levels; for EIT, this value also drifts up after 2003, maybe due to detector degradation. There are high frequency fluctuations in this ratio for each of the instruments, which may reflect frame-to-frame changes. For this study focusing on relatively long-lived coronal holes that persist for many days, we consider the short-term fluctuations unimportant. These fluctuations introduce subsequent fluctuations in the calculation of coronal hole area and magnetic fluxes, which will be addressed later in the paper.

\begin{figure*}[h!tbp]
\centering
\includegraphics[width=18cm]{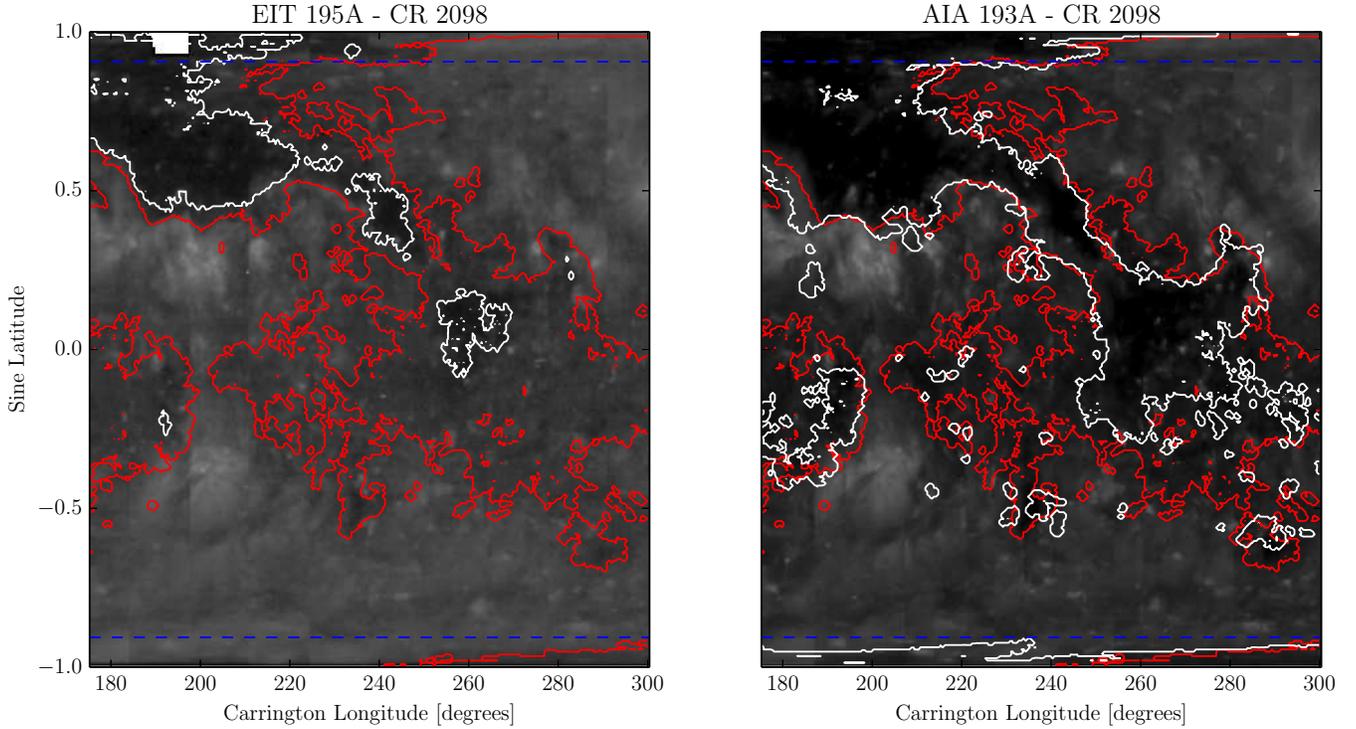}
\caption{Left: EUV synoptic image from EIT 195 \AA, with contours of calculated coronal hole boundaries from EIT data in white and AIA data in red. Right: EUV synoptic image from AIA 193 \AA, with contours of calculated coronal hole boundaries from AIA data in red, and EUVI A/B data in white. The enhanced contrast of the AIA instrument is immediately apparent for the calculation of coronal hole boundaries.}
\label{fig:f5}
\end{figure*}

Figure~\ref{fig:f5} shows the coronal hole boundaries detected in the EUV images obtained by different instruments, using this automated thresholding technique. The left panel displays a synoptic EUV image gathered from EIT 195 \AA, with contours displaying the calculated coronal hole boundaries from EIT data (white) and AIA data alone (red). The right panel displays a synoptic EUV image gathered from AIA 193 \AA, with contours of coronal hole boundaries from AIA data alone (red) and EUVI A/B data alone (white). The comparison of the two images as well as the two contours show evidently that the superior contrast of AIA and EUVI images allow better detection of coronal holes. The polar hole is captured in both images, but the AIA/EUVI polar hole is larger than the EIT hole. Smaller or weaker holes at lower latitudes are evidently present and detected in the AIA/EUVI image, but hardly seen in the EIT image. Therefore, AIA and EUVI measurements are crucial to study coronal holes in middle-low latitudes. As these holes may be located in stronger magnetic fields compared with polar holes, they may contribute significantly to the magnetic open flux. This will be further discussed in Section 4 when open flux measurements are compared.

\subsection{Coronal hole determination}
\begin{figure*}[h!tbp]
\centering
\includegraphics[width=18cm]{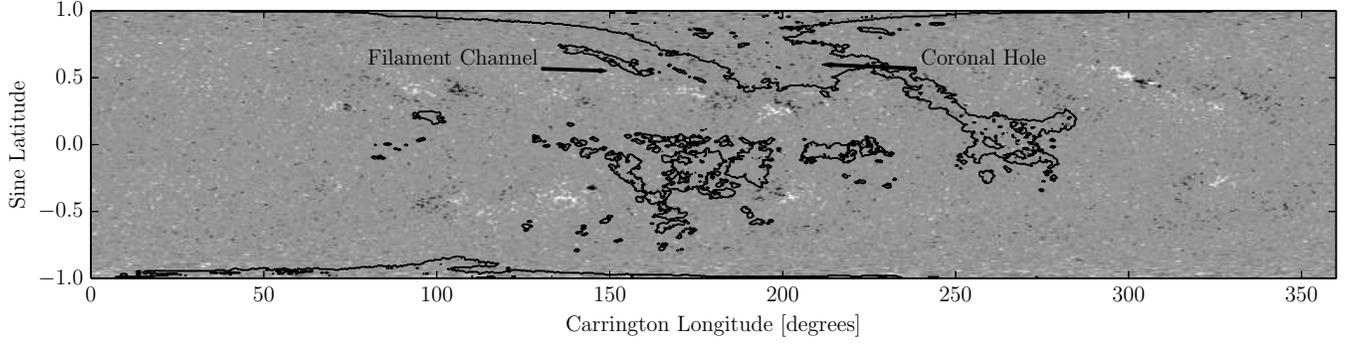}
\caption{HMI synoptic map of radial magnetic field for Carrington rotation 2098, with contours of initially suspected coronal hole regions. Each source of data, AIA and EUVI A/B, were run through the thresholding routine, producing a DN threshold value for each source. Contours were produced of regions below this thresholding value, and projected into a Carrington equal area projection.}
\label{fig:f6}
\end{figure*}

The thresholding technique is able to identify dark features in the EUV images, including both coronal holes and filaments or filament channels. To automatically distinguish coronal holes from filaments, we also incorporate magnetic field information from synoptic charts of radial magnetic flux. Coronal hole regions are most often clearly dominated by a single magnetic polarity over their entire area \citep{2009SSRv..144..383W}. Filament channels are characterized by depleted intensity in EUV wavelengths, similar to coronal holes. However, filament channels lie along a polarity inversion line, and thus lie atop a region which will tend to be magnetically neutral as a whole. Therefore, in addition to the criteria of EUV intensity, a coronal hole candidate region must also tend towards a dominating polarity. With the thresholding on each instrument completed for a particular time frame, each array is converted into a boolean array, with 0 indicating regions above the threshold value for that instrument at that particular time. A value of 1 in a pixel marks areas below the threshold value. These individual arrays are then projected into a Carrington equal area map, marked in longitude and sine latitude. Each of the suspect regions are overlaid with a synoptic map of radial magnetic field, as displayed in Figure~\ref{fig:f6}. For each region in question composed of $N$ pixels, there exist $N$ values of magnetic flux, defined as $\Phi$. We can define the skew of the magnetic flux as, denoting it as the variable $\gamma$,

\begin{equation}
	\gamma = \frac{1}{N} \sum_{j=0}^{N-1} \left( \frac{\Phi_j - \bar{\Phi}}{\sigma} \right)^3
\end{equation}

Here $\bar{\Phi}$ denotes the mean flux value within a particular region and $\sigma$ denotes the standard deviation of the flux values within the region. The value of magnetic flux skew is calculated for each suspected coronal hole region in this way. Regions whose magnetic flux skew exceed 0.5 in magnitude are labeled as coronal holes, the criteria for selection. This value was chosen through a manual search of select regions, finding this value to accurately distinguish between the two phenomena.

\begin{figure}[h!tbp]
\centering
\includegraphics[width=7.5cm]{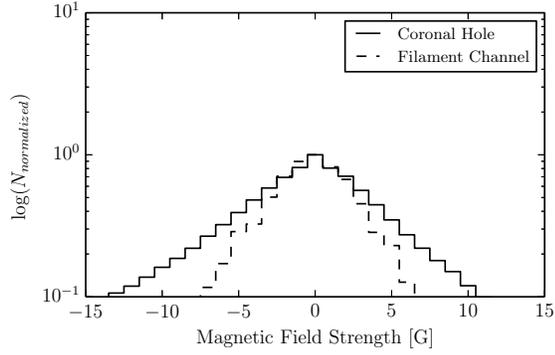}
\caption{Example of the magnetic flux distribution within two suspected coronal hole regions. The solid line displays a histogram of magnetic flux for a positively identified coronal hole region. The magnetic flux within the coronal hole boundary is slightly unbalanced, tending toward negative polarity. The dashed curve is data from an identified filament channel, which is more balanced in flux distribution. Note that the distributions vary outside of the HMI noise level, $\pm$ 2.3 G.}
\label{fig:f7}
\end{figure}

Magnetic flux histograms are compared for the labeled coronal hole and filament channel in Figure~\ref{fig:f7}. The histogram of magnetic flux in the filament channel shows a relative balance between positive and negative magnetic flux. The histogram for the magnetic flux in the coronal hole, however, is noticeably skewed towards negative magnetic flux values.

\begin{figure}[h!tbp]
\centering
\includegraphics[width=7.5cm]{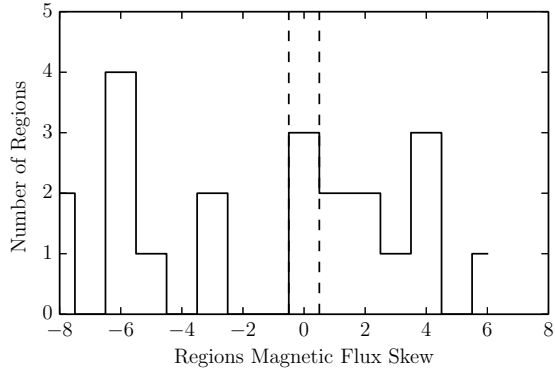}
\caption{Histogram of the calculated magnetic flux skew values for the particular time frame under consideration. Vertical dashed lines indicate the magnetic flux skew cutoff value, 0.5. Regions whose calculated magnetic flux skew are less than 0.5 in magnitude are identified as filament channels.}
\label{fig:f8}
\end{figure}

Figure~\ref{fig:f8} displays a histogram of the magnetic flux skew values for each of the suspected coronal hole regions. Vertical dashed lines display the cutoff value of $\pm 0.5$. For EIT measurements, the MDI synoptic charts of radial magnetic field flux are used; and for AIA-EUVI measurements, the HMI synoptic charts of radial magnetic field flux are used. 

\subsection{Persistence, area, and open flux of coronal holes}
EUV images are analyzed with the above-mentioned technique to determine coronal hole boundaries. For a given dataset, each individual image frame is characterized by a discretized set of spatial and temporal coordinates, a set of latitudes, longitudes, and time frames, $\{\theta, \phi, t\}$. This set is discretized such that each is indexed by an integer value, $\{\theta_i, \phi_j, t_k\}$. For each of these values, a byte value is assigned that indicates the presence of a coronal hole, defined as a scalar function, $\psi(\theta_i, \phi_j, t_k)$. For a particular range of time, a `persistence map' can be defined as the overall persistence of a coronal hole within a particular location. This map can be defined as,

\begin{equation}
\Psi(\theta_i, \phi_j) = \sum_{k} \psi(\theta_i, \phi_j, t_k)
\end{equation}

Note that this persistency is not continuous in nature, but totaled over the entire period under consideration. Nevertheless, the persistence map illustrates the overall distribution pattern and persistence of coronal holes, which, as a global property, can be compared for different instruments or models.

In considering the persistency array data, it is crucial to note that though the view from each instrument overlaps, the persistency data does not over-count pixels. After stitching the view from each instrument together, the coronal hole map array is flattened into a binary map, only noting the existence of a coronal hole without regard to the number of instruments that may have observed it. Coronal hole surface area is calculated by summing for each particular frame in the persistence array for the instrument set under consideration,

\begin{equation}
A_{CH}(t_k) = R_\odot^2 \sum_{i,j} \psi(\theta_i, \phi_j, t_k) \cos(\theta_i)\end{equation}

In addition to the spatial hole data, magnetic flux measurements are available. Using the persistence information previously mentioned, the magnetic flux can be defined via, 

\begin{equation}
\Phi_M(t_k) = R_\odot^2 \sum_{i,j} \psi(\theta_i, \phi_j, t_k) \cos(\theta_i) B_r(\theta_i, \phi_j, t_k)
\end{equation}

Here, these flux measurements are obtained using calibrated synoptic charts of radial magnetic field strength. For coronal holes observed with EIT data from 1996 through 2011, the MDI synoptic charts are employed. We sum the radial magnetic field in the coronal holes with a cutoff magnetic field strength at $\pm$ 5.0 G, which was determined to be the noise level of the MDI synoptic charts \citep{2012SoPh..279..295L}. For coronal holes measured with AIA-EUVI data from 2010 June to 2013 January, the HMI synoptic charts are used with a cutoff field strength at $\pm$ 2.3 G, the noise level for the HMI synoptic charts calculated by \citet{2012SoPh..279..295L}. 

Table~\ref{table:table2} displays the standard deviation values for $A_{CH}$ (surface area) and $\Phi_M$ (magnetic flux), and in each latitude region under consideration for each instrument / calculation set. Full, northern pole (NP), low-latitudes (LL), and southern pole (SP) regions are defined from latitudes of -90:90, 65:90, -65:65, and -90:-65 degrees, respectively. These values are calculated from the time series of each quantity, and represent the uncertainty in their calculation stemming from the variation in calculated thresholding value.

\begin{deluxetable}{rrrr}
\tablecolumns{4}
\tablewidth{0pt}
\tablecaption{Uncertainty in coronal hole / open field quantities}
\tablehead{ 
\colhead{Dataset} & \colhead{Region} & \colhead{$\sigma_{A_{CH}}~[10^{20} cm^2]$} & \colhead{$\sigma_{\Phi_M}~[10^{21} Mx]$}}
\startdata 
EIT & Full & 13.7 & 12.4\\
	& NP & 8.51 & 5.59\\
	& LL & 8.62 & 9.30\\
	& SP & 7.28 & 5.71\\
PFSS & Full & 13.8 & 7.10\\
	& NP & 3.80 & 2.81\\
	& LL & 13.8 & 7.51\\
	& SP & 3.03 & 2.28\\
AIA/EUVI & Full & 9.02 & 5.38\\
	& NP & 0.432 & 0.319\\
	& LL & 7.79 & 3.95\\
	& SP & 2.47 & 2.31
\enddata 
\label{table:table2}
\end{deluxetable}

Whereas full surface map may help capture more coronal holes, one limitation is the partial magnetic field availability. The line of sight magnetic field measurements are only taken onboard the SDO spacecraft, and are therefore only for one vantage angle. The effects of possible changing magnetic fields on the back-side will be examined in the ensuing study. In this present study, we assume that for the course of one rotation the relevant magnetic field does not drastically change in the coronal hole interior regions, therefore a synoptic chart approach may be taken. 

In the following section, we will compare the area and open flux of coronal holes and their latitude dependence using EIT and AIA-EUVI observations, as well as from the PFSS model.

\section{Properties of Persistent Coronal Holes}

\subsection{EIT Observations}

We have analyzed the entire EIT dataset in 195\AA. This dataset spans Carrington rotation 1920-2105, corresponding to the time period from 1996 May 31 to 2010 December 31, as referenced in Table~\ref{table:table1}. This dataset provides a unique opportunity to consistently study coronal holes throughout the activity variation along a solar cycle. The EIT data are analyzed at the cadence of one image per day; coronal holes detected with this cadence are rather long-lived features, or persistent coronal holes. Our analysis shows that, as one would expect, the majority of coronal hole persistence occurs near the polar regions above 60 degrees in latitude. Over one solar cycle from the solar minimum (1996 - 1997) to solar maximum (2000 - 2003), coronal holes gradually extend downward from the polar cap to lower latitudes. These variations can be better characterized by the evolution of the coronal hole area and open flux at varying latitudes.

\begin{figure*}[h!tbp]
\centering
\includegraphics[width=18cm]{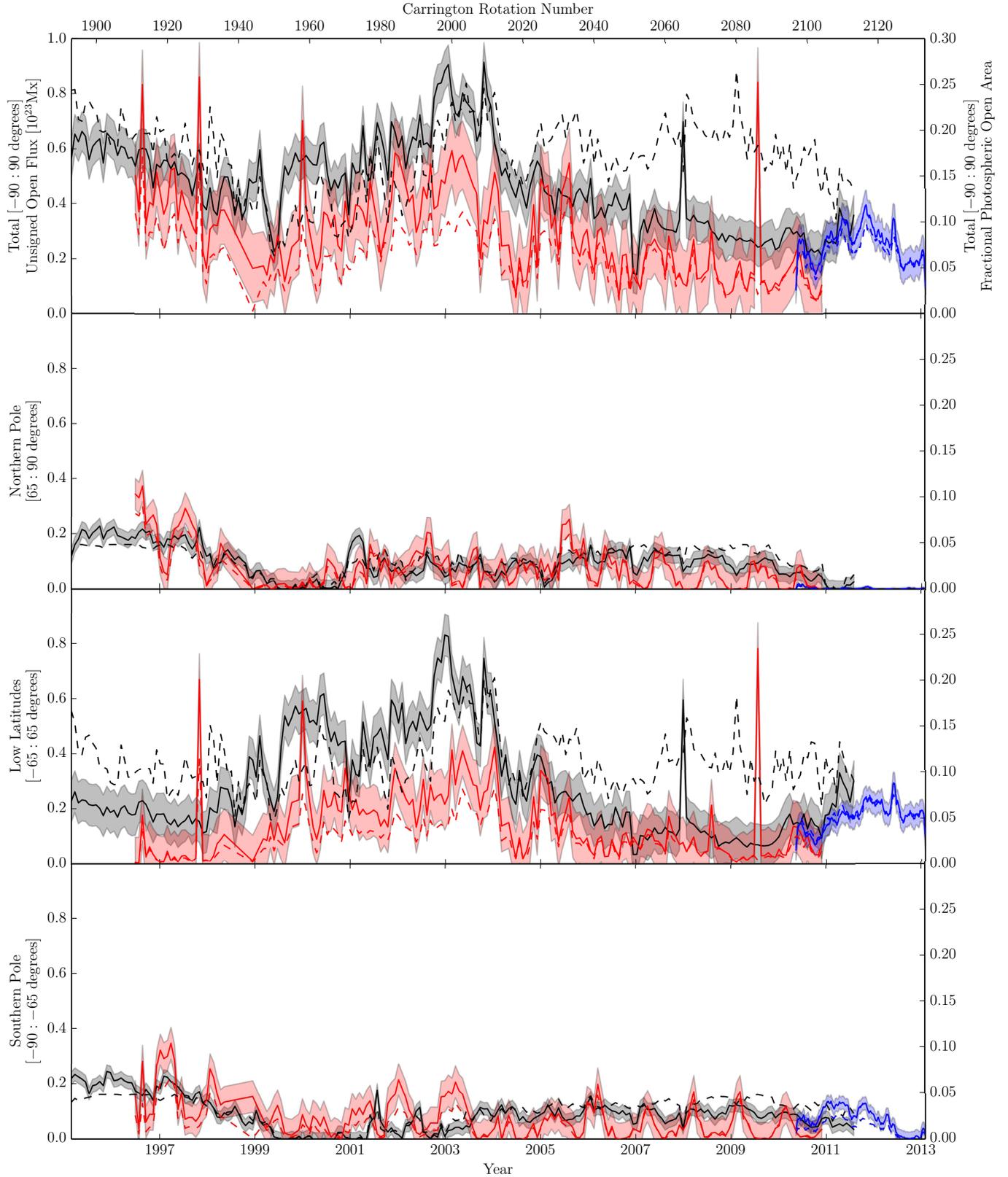}
\caption{Unsigned open magnetic flux and fractional photospheric open area for calculated PFSS field, as well as observed coronal hole boundaries from EIT and AIA/EUVI. Unsigned open magnetic flux and open area are displayed with solid and dashed curves, respectively. PFSS, EIT, and AIA/EUVI results are color coded as black, red, and blue, respectively, with corresponding shaded regions displaying the standard deviation of the flux values. The top panel displays the open magnetic flux and areas for the entire photospheric surface, while the bottom three panels display the northern pole [90:65 degrees], low latitudes [65:-65 degrees], and southern pole [-65:90 degrees].}
\label{fig:f9}
\end{figure*}

Figure~\ref{fig:f9} displays the coronal hole area, as well as the total unsigned magnetic flux determined from these coronal hole boundaries by summing the radial magnetic field measured from the synoptic charts of MDI. To further distinguish contributions to the open flux from different latitudes, we also plot the areas and fluxes from the north polar holes (defined for latitudes from 65 to 90 degrees), the middle-low latitude region (defined from -65 to 65 degrees), and south polar holes (from -90 to -65 degrees), respectively. These are compared with the open flux and area computed using the PFSS model, as will be described later in the text.

The top panel of the figure shows the total area and total open magnetic flux, which are roughly correlated during the solar cycle. The total surface area measured from EIT (red dashed line) reaches more than 15\% of the total solar surface area during the minimum between 1996 and 1997, and then rapidly decreases until 1999; after the solar maximum, the coronal hole area increases again to about 10\% in 2003, and then gradually decreases to nearly 5\% during the last extended minimum between 2006 - 2010. Correspondingly, the total open flux rapidly decreases from 5$\times 10^{22}$ Mx (if we ignore the few spikes) in mid-1996, to the minimum of 2$\times 10^{22}$ Mx in 1999, and then increases to be around 5$\times 10^{22}$ Mx in 2003; the total open flux stays low at around 2$\times 10^{22}$ Mx after 2006. The total open flux measured from EIT coronal holes vary by a factor of 2-3 from minimum to maximum.

The lower three panels show the latitude-dependence of the open flux. It is evident that contributions from the polar holes and middle-low latitudes vary in different ways during the solar cycle. Significant polar holes in both the north and south are captured in EIT observations during the minimum from 1996 - 1998, which then shrink until 1999, and then grow again afterwards. During the last minimum, the total polar hole area is relatively low covering less than 5\% of the total solar surface. The open flux from the polar holes is strictly correlated with the hole area, reflecting the relatively smooth magnetic field distribution in polar holes. The middle-low latitude holes have the largest area coverage shortly after the solar maximum; from 2000 to 2004, these holes cover an area between 5-10\% of total solar surface, and contribute to the total open flux more significantly because of the stronger magnetic field in these regions. During the solar minima, the total hole area is rather low at 1-2\% on average. Note that there are larger holes in the middle-low latitudes during the past solar minimum between 2006 - 2008 than the previous one between 1996 - 1998.

Measurements of polar hole area and open flux were also conducted by \citet{2002SoPh..211...31H} using observations at the chromospheric He I 10830 line from National Solar Observatory / Kitt Peak (NSO/KP). These measurements cover the period from 1990 - 2000, partly overlapping with the EIT measurements in the present study. Comparing these two measurements during 1996 - 2000, it is seen that the annual variations in the two measurements are quite consistent for both the north polar holes and south polar holes. Both measurements show that the north polar holes cover up to 8\% surface area and south polar holes cover up to 7\% surface area between 1996 and 1997, followed by a decrease in area. Both measurements show, in both polar holes, a local minimum around 1997, a peak in middle 1997, another minimum around 1998, and then a peak afterwards. From 1998 to 2000, the north polar hole diminishes faster than the south polar hole, as shown in both observations. In terms of the open flux, \citet{2002SoPh..211...31H} used NSO/KP magnetograms, with MDI magnetograms used in the present study. Both measurements yield the maximum open flux of 3$\times 10^{22}$ Mx in either north or south polar holes between 1996 and 1998, which decreases in the same fashion as the area decreases. Therefore, the comparison between these two independent studies during this overlapping time period shows very good consistency in measuring polar holes using two different kinds of observations.

\subsection{AIA-EUVI Observations}

The EIT instrument onboard SoHO is limited to the view from the Earth-Sun line; furthermore, we have shown that EIT observations underestimate the area of low-latitude coronal holes, which we attribute to a higher level of instrument scattered light. AIA and EUVI provide a comparable dataset to that used on EIT, with higher sensitivity to coronal holes outside the polar caps, and with the added bonus of multiple viewpoints. AIA-EUVI data are obtained from Carrington rotations 2096-2133. This spans the time period from 2010 May to 2013 January, as referenced in Table~\ref{table:table1}. We analyze these data sampled at 12 hour cadence to capture persistent coronal holes.

From these data, we measure the coronal hole area and open flux, displayed by dashed and solid blue lines, respectively, in Figure~\ref{fig:f9}. The top panel shows the total surface area and total unsigned flux. Coronal holes cover the total solar surface by 5-10\% from 2010 - 2013, and the total flux in these holes is measured to range from 2$\times 10^{22}$ Mx in late 2010 and late 2012 to 4$\times 10^{22}$ Mx around 2012. The total flux is entirely correlated with the hole area, and varies by a factor of 2.

The latitude dependence of the open flux is examined in the following panels, showing that the north polar holes are present only for a short period during mid-2010, having diminished since then. The south polar holes persist for two more years and diminish in late 2012. These holes contribute to up to 1.7$\times 10^{22}$ Mx open magnetic flux. Coronal holes in the middle-low latitudes contribute significantly, and account for twice as much open flux as from the south polar holes. After mid-2012, middle-low latitude holes over-dominate.

It is noted that, during the period of the second half of 2010 when both EIT and AIA-EUVI observations are analyzed, the hole area measured from AIA-EUVI observations is significantly greater than that measured in EIT observations. The figure suggests that, in both the south polar regions and middle-low latitude regions, AIA-EUVI sees larger holes than EIT, while in the north polar region, EIT holes cover a larger area than AIA-EUVI during the brief period of mid-2010. These differences lead to an overall greater open flux measured with AIA-EUVI observations during the overlapping period. We note that the open flux in EIT holes is measured using synoptic MDI magnetograms, and the open flux in the AIA-EUVI holes is measured using synoptic HMI magnetograms. Nevertheless, using different magnetograms does not change the total flux measured in the holes; the difference in the flux measurements is solely produced by the difference in the hole area measurements. These discrepancies will be further discussed in the next section, and also with respect to their comparison with model results.

\subsection{PFSS Model Comparison}

The widely used potential field source surface (PFSS) model, following the work of \citet{1992ApJ...392..310W}, is employed for this study to reconstruct the global coronal magnetic field. The PFSS model is an extrapolation method that assumes a potential field below a nominal source surface set at 2.5 solar radii \citep{1969SoPh....6..442S, 1984PhDT.........5H}, with field lines above 2.5 solar radii being open. The model uses monthly synoptic charts composed from photospheric magnetograms as the boundary condition to compute the potential field. This implies an underlying assumption that the large-scale magnetic field is static over one rotation. These assumptions ignore evolution of the photospheric magnetic field and the possibility that the solar corona is not entirely current free. To circumvent some of these problems, \citet{2003SoPh..212..165S} used a few techniques to improve the boundary condition; they used a flux dispersal model to evolve the photospheric magnetic field, applied a data assimilation technique to update the synoptic chart by inserting newly emerging active regions on the disk, and inferred active region locations on the back side of the Sun with helioseismology approaches. Variations of the PFSS reconstruction methods and numerous MHD models have also been developed with their results compared against one another. These models are reviewed by \citet{2012LRSP....9....6M}. In this study, we focus on a first-order comparison of global patterns of persistent coronal holes, so the standard PFSS reconstruction is adopted despite various issues related to this model. In addition, the PFSS model has been widely used to calculate open magnetic flux, which is then compared with various other measurements or calculations by other models. Therefore, PFSS results can provide a baseline reference to compare our measurements to various other studies.

We compute the spherical harmonic coefficients from the longitudinal magnetograms obtained by Wilcox Solar Observatory to 29 orders, and use them to reconstruct the magnetic field at the solar surface as the boundary condition. For potential magnetic field, a magnetic scalar potential is defined which satisfies $\nabla^2 \Phi_M = 0$. And the potential magnetic field below the source surface $2.5R_\odot$ is then calculated as, $\mathbf{B} = - \nabla \Phi_M$.

A field-line tracking routine is then used to map the foot-prints of open field lines to the solar surface, namely the model computed coronal holes. Open field area and magnetic flux values are calculated at the photosphere in a manner similar to EIT and AIA/EUVI coronal hole observations, using the open field regions as a mask in this situation. The model computed coronal hole area and open flux are then compared with observations, as presented in Figure~\ref{fig:f9}. In the figure, the model computed coronal hole area and open flux are plotted in dashed and solid black lines, respectively.

First, comparing EIT observations with the PFSS results, it is seen that the solar cycle variation of the total open flux is present in both measurements, but the total hole area and open flux from EIT measurements are consistently smaller than derived from the PFSS model throughout the solar cycle. The EIT flux is closest to the PFSS flux during the last solar minimum until 1998, with $\Phi_{EIT} \approx 0.8 \Phi_{PFSS}$. During the solar maximum, the EIT measured flux is only about 60-70\% of the PFSS flux. The EIT flux temporarily matches the PFSS flux between 2005 and 2006, and then drops to about 50\% of the PFSS flux since then. It also appears that, from 2005 to 2011, the total hole area measured using the PFSS model is significantly higher than measured in EIT data, when the correlation between hole area and total flux is not preserved in the PFSS measurements during this period.

A more detailed comparison of latitude-dependence shows that the model and observation are relatively consistent in measurements of polar holes. There is very good agreement between modeled and observed hole area as well as open flux in both the north and south polar caps during the previous solar minimum (1997 - 2001); during the maximum (2001 - 2005), the PFSS model and observation agree with each other for north polar holes, but do not compare very well in south polar holes. From 2005 - 2011, which covers an extended solar minimum, the model and observational measurements in both polar holes show the same temporal variations in the open flux, which roughly exhibit a one-year period caused by the solar P-angle variations, yet fluctuations in the EIT observation have greater amplitudes. During this period, the PFSS computed open surface area is significantly larger even in polar holes.

There is marked disagreement in middle-low latitude holes between model and EIT observations. During the solar maximum, the model computed hole area and open flux are systematically larger than EIT measurements by nearly a factor of two. During the minima, the PFSS model appears to produce significantly more coronal hole area than observed by EIT, even though the calculated open flux is not as much due to weak magnetic field during the solar minima. It is likely that the PFSS model uses a very coarse resolution, which may artificially amplify open field regions. Other studies have indicated that PFSS tends to over-estimate the steady-state open flux \citep{2008ApJ...674.1158L}.

Some insight can be gained by also comparing with the AIA-EUVI measurements plotted in blue lines in Figure~\ref{fig:f9}. For one and one-half years from 2010 June to 2012 January, the AIA-EUVI measured total open flux appears to match the PFSS open flux in both the evolution trend and absolute amount. Looking into details, in the middle-low latitudes, although the AIA-EUVI open flux more closely matches with the model flux than EIT does, it is still smaller than the model flux, and the PFSS hole area is still much larger than observed.

\begin{figure*}[h!tbp]
\centering
\includegraphics[width=18cm]{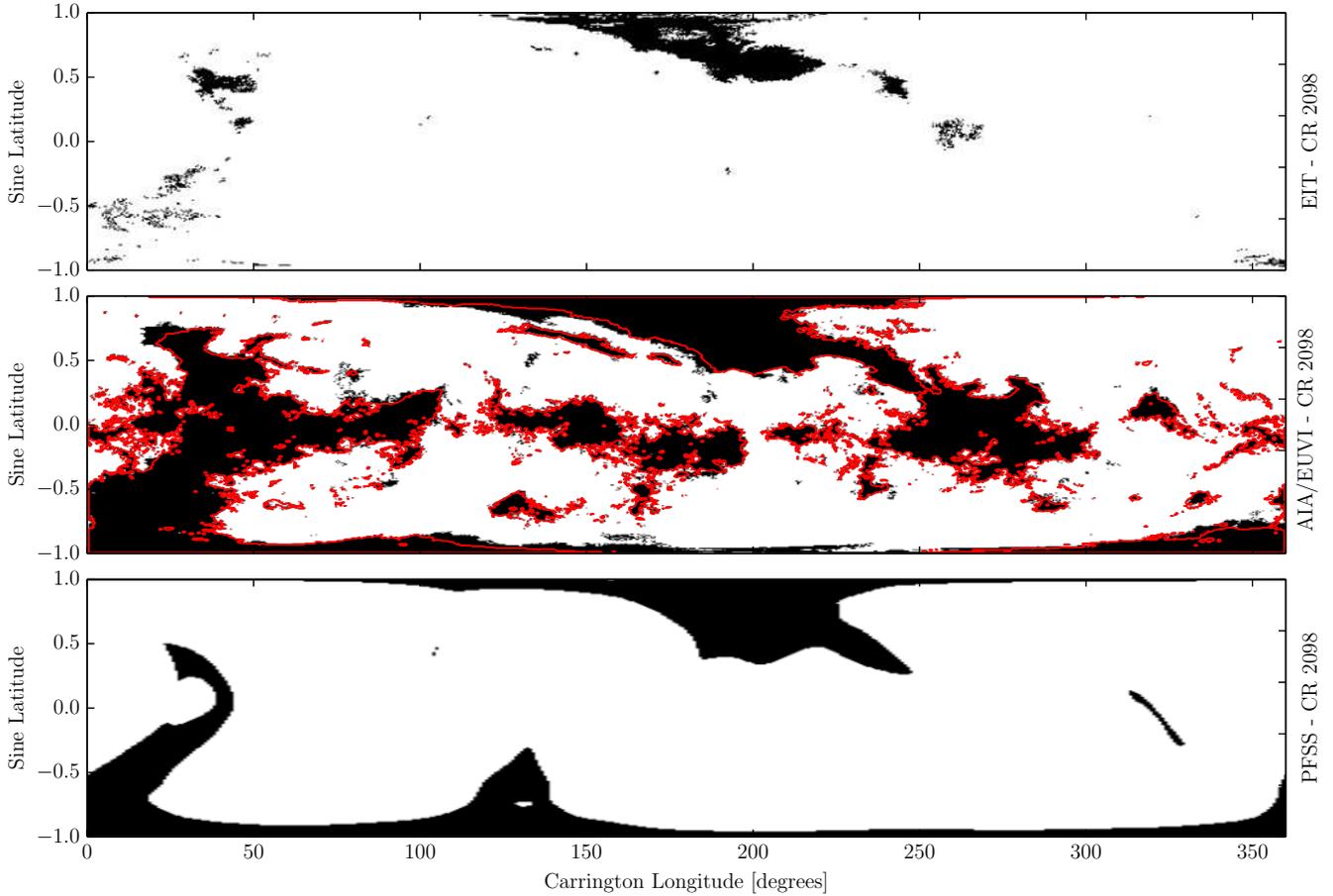}
\caption{Coronal hole and open magnetic field boundaries for Carrington rotation 2098. The upper panel displays the observed coronal hole boundaries from EIT. The middle panel displays the observed coronal hole boundaries from AIA/EUVI in black, with a contour of AIA observations alone in red. The lower panel displays the open magnetic field boundaries from our PFSS calculation.}
\label{fig:f10}
\end{figure*}

As a short summary, these plots show that the PFSS model tends to produce larger hole areas during most of the solar cycle, as compared with EIT and/or AIA-EUVI observations. The model also produces more open flux primarily in middle-low latitudes than exhibited in both observations. To understand these discrepancies, Figure~\ref{fig:f10} shows, as an example, the comparison of coronal hole boundaries in three different measurements during CR 2098 in 2010 June.

Seen in this comparison, all three measurements yield comparable hole boundaries in polar regions defined at greater than 65 degrees, excusing poor south polar coverage from EIT. It is evident that large discrepancies are present at middle-low latitudes. Compared with AIA-EUVI, EIT clearly cannot detect coronal holes to the same degree in this region, which explains the overall flux deficiency measured in EIT holes, particularly during the solar maximum, when low-latitude holes contribute more significantly. The comparison between PFSS holes and AIA-EUVI holes also shows that PFSS tends to over-estimate the areas of holes in the middle latitude region, which appear to extend from polar holes - these were called polar hole lobes by \citet{2002SoPh..211...31H}. This is most evident at 130 degrees longitude in the southern pole and 210 degrees longitude in the northern pole. On the other hand, a series of smaller coronal holes at low-latitudes detected in AIA-EUVI images are missed by PFSS for this rotation. These discrepancies in middle and low latitude regions between PFSS and AIA-EUVI explain why the two measurements seem to yield a comparable amount of total open flux yet PFSS measures much larger hole areas than AIA-EUVI.

\begin{figure*}[h!tbp]
\centering
\includegraphics[width=18cm]{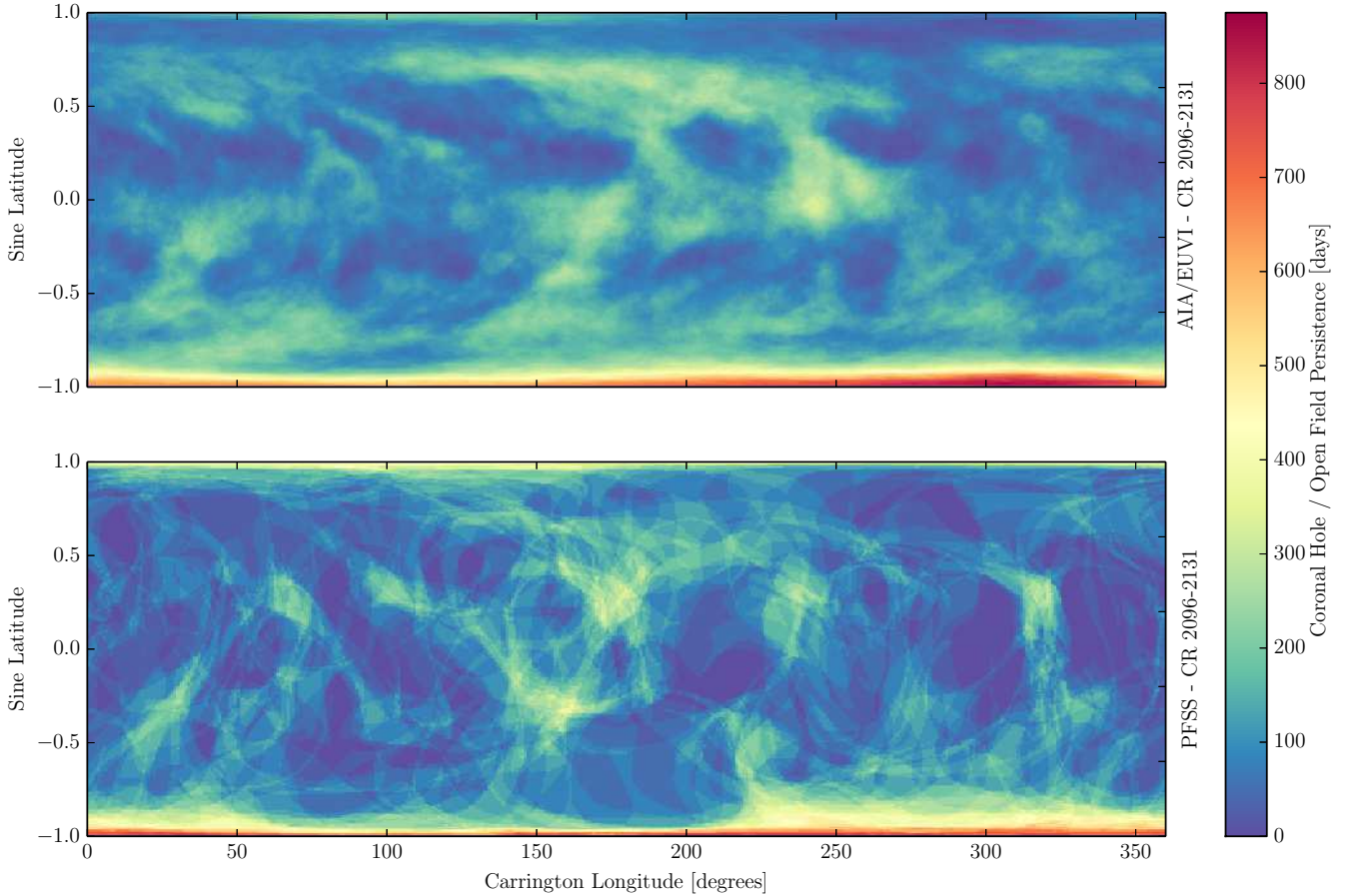}
\caption{Coronal hole persistence map for a combination of AIA 193\AA~and EUVI 195\AA~datasets, along with a corresponding map generated using spherical harmonic coefficients obtained from the Wilcox Solar Observatory and a PFSS open field reconstruction. Persistence is scaled in days of non-consecutive persistence of coronal hole / open field for each pixel.}
\label{fig:f11}
\end{figure*}

For a more complete comparison beyond one rotation, Figure~\ref{fig:f11} displays the observed and computed persistence maps for Carrington rotations 2096-2131, from 2010 May to 2013 January. The reconstructed open field map matches with observations only in very general global patterns. Both the reconstruction and observations show that coronal holes in the southern pole are the most persistent feature throughout this period. Discrepancies are present concerning the relative hole coverage at different latitudes. These persistency maps will aide studies that track the spatial and temporal evolution of coronal holes \citep{1975SoPh...42..135T, 2001ApJ...560L.193M, 2002GeoRL..29.1290M}. However, the detailed study of evolution of coronal holes is beyond the scope of the present paper, and will be conducted in the near future.

\begin{figure}[h!tbp]
\centering
\includegraphics[width=7.5cm]{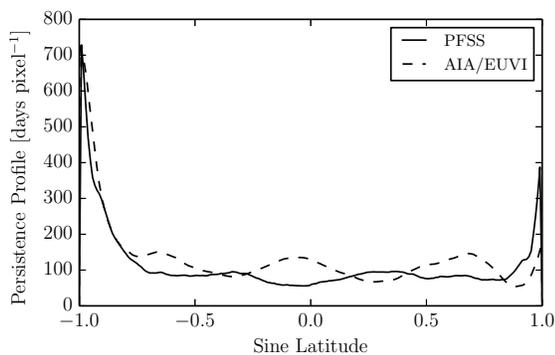}
\caption{Coronal hole persistence map projected as a function of latitude. The two profiles are in units of $days$ $pixel^{-1}$. The PFSS model calculates a further extension of open field persistence from the poles as compared with our AIA/EUVI observations. The persistence value at low latitudes agrees in value, but with variations in distribution visible in Figure~\ref{fig:f11}.}
\label{fig:f12}
\end{figure}

To further quantify the comparison, the persistence map is compressed along longitudes to yield a profile for coronal hole persistency as a function of latitude,

\begin{equation}
\lambda (\theta_i) = \sum_j \Psi(\theta_i, \phi_j)
\end{equation}

Figure~\ref{fig:f12} illustrates this projection, reflecting the latitude dependence of the coronal hole coverage obtained by the two methods. Again, the plot shows similar global patterns with discrepancy in the relative hole distribution along the lower-latitudes. For the analyzed period, the PFSS reconstruction appears to have slightly over-estimated the northern polar hole area and the extension to middle latitudes. Caution should be taken, though, that the persistency in units of days in these maps is not directly proportional to the measured open flux, since the hole open flux are measured once per rotation for the potential field model.

\subsection{Reversal of Magnetic Open Flux}

\begin{figure*}[h!tbp]
\centering
\includegraphics[width=18cm]{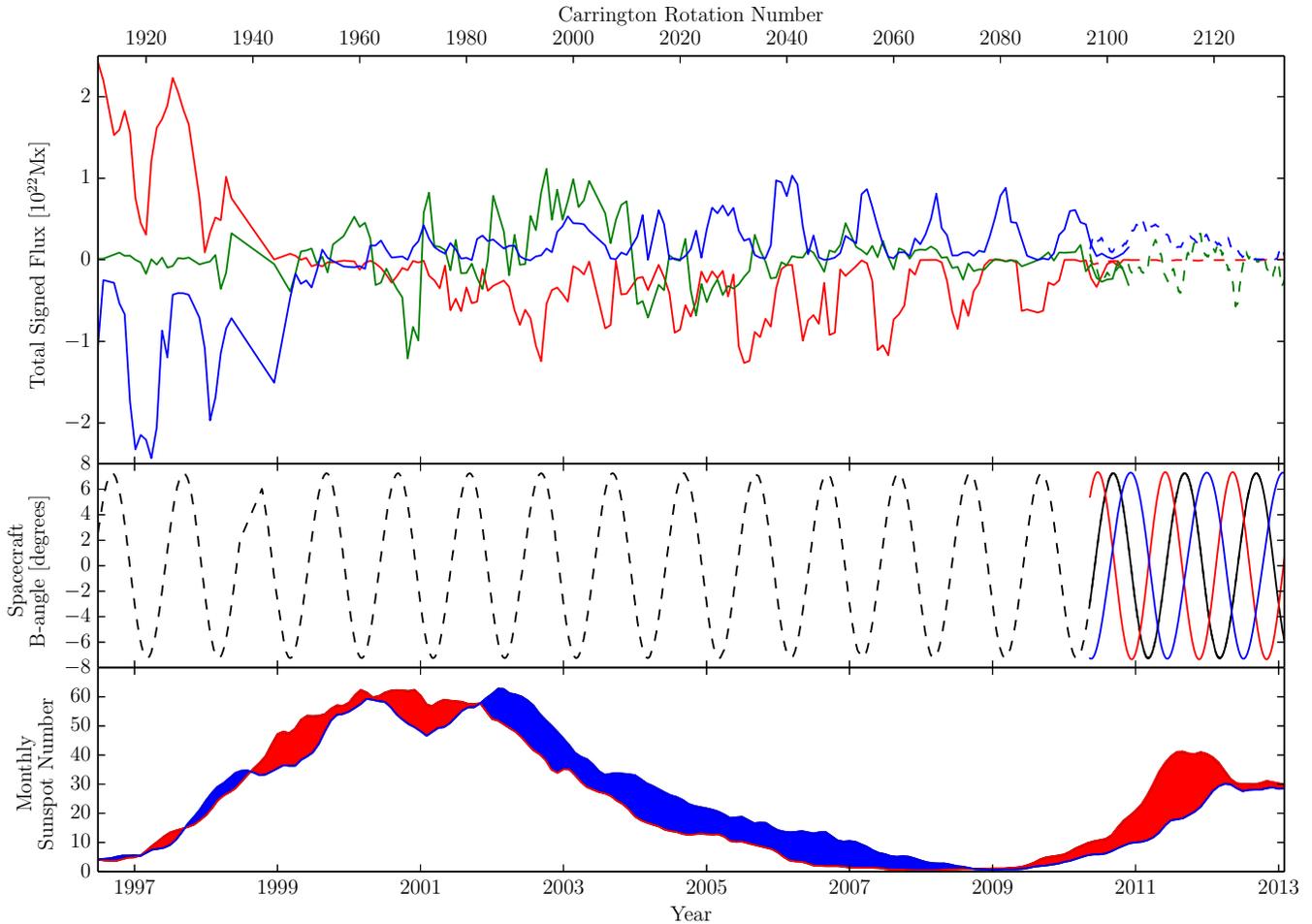}
\caption{(top) Signed open magnetic flux from EIT/AIA-EUVI surface maps of coronal hole boundaries. Flux is subdivided into three segments depending on location. The southern polar, low-latitude, and northern polar regions extend between [-90,-65], [-65,65], [65,90] degrees latitude, and are displayed in blue, green, and red lines respectively. EIT-era observations are marked with a solid line, with AIA/EUVI observations marked with dashes. (middle) Spacecraft B-angle for the various instruments employed in the upper figure. SoHO, SDO, STEREO-A, and STEREO-B are displayed in dashed black, solid black, red, and blue, respectively. (bottom) Smoothed monthly sunspot number for the northern and southern hemisphere in red and blue, respectively. The area between curves is color-coded to indicate which hemisphere dominates. Sunspot data via WDC-SILSO, Royal Observatory of Belgium, Brussels.}
\label{fig:f13}
\end{figure*}

Apart from the total unsigned flux from coronal holes, the total signed magnetic flux is also measured in coronal holes at different latitudes, namely the polar holes and middle-low latitude holes, as defined previously. Figure~\ref{fig:f13} displays signed magnetic flux in coronal holes measured from 1996 to 2013, first by EIT, and then by AIA-EUVI as well. Note that no correction has been made here for relative spacecraft B-angle, as is most apparent from the variation in polar data availability from the EIT dataset. Without making assumptions as to polar magnetic field data or coronal hole coverage, the overall trend is still visible.

Here, the temporal evolution of the flux reversal is evident. Whereas from 1996 until 2000 the northern pole is dominated by positive flux, and the southern pole dominated by negative flux, this trend reverses after 2000. Moreover, the overall magnitude of the signed flux indicates that the relative dominance of the respective polarities is much weaker after this reversal. Observations from AIA-EUVI maps have been displayed in dashed style. These values match well during the overlap with EIT.

During the period observed by AIA/EUVI, the southern pole is dominated by strong positive flux, the northern pole is weakly dominated by negative flux, and the low latitudes are more mixed, with a slight tendency towards negative fluxes. Here, the time evolution is more evident, as the low latitude regions are mostly negative for the first nine months of observation, where they transition to positive dominance. This then reverses back and forth a few times. 

\section{Conclusions and Discussion}

In this study, an automated method was devised and applied to detecting coronal hole boundaries using full-disk EUV imaging observations from a combination of instruments. SoHO:EIT, SDO:AIA, and STEREO:EUVI A/B data were considered over the respective instrument lifetimes. SoHO:EIT data was collected from 1996 through 2011, covering one and one-half solar cycles. The combined observations from SDO:AIA and STEREO:EUVI A/B provide nearly full-surface coverage, but over a shorter timespan from 2010 May to 2013 January. This combination of data provides a unique opportunity for continuous tracking of long-lived or persistent coronal holes. Aided with observations of photospheric magnetic field acquired by MDI and HMI, the magnetic flux is measured within defined coronal hole boundaries. Assuming that the magnetic flux within coronal hole boundaries largely represents magnetic open flux, we compare these results with a potential field source surface model.

The area, magnetic flux, and their latitude dependence of coronal holes are measured using EIT images from 1996 through 2011. The measured total area of coronal holes varies between 15\% and less than 5\% of total solar surface area, and the total unsigned open flux varies from 5$\times 10^{22}$ Mx (1996-1997), to 2$\times 10^{22}$ Mx (2000), back to 5$\times 10^{22}$ Mx (2003), and then 2$\times 10^{22}$ Mx since 2006. There is a good agreement between these results and the previous study by \citet{2002SoPh..211...31H} of polar holes using He I 10830 data during the period 1996 through 2000, when both data are analyzed. The two studies yield comparable measurements of polar hole area, open flux, and their annual variations. When compared with the PFSS model, it is shown that the model and EIT measurements exhibit similar global patterns of coronal hole coverage particularly in high-latitude ($>65$ degrees) polar regions, as well as similar solar cycle dependence of open flux consistent with model results by \citet{2009SSRv..144..383W} and IMF measurements by \citet{2010JGRA..115.9112Y}. However, the model yields larger area and more open flux than observed, particularly in lower latitude areas and during the solar maximum. There is also a large discrepancy in both area and flux during the extended past solar minimum from 2006 - 2011, even though the two measurements show the same evolution trend in general.

Measurements are made with AIA-EUVI observations from 2010 May through 2013 January, and compared with EIT observations as well as the PFSS model results. It is found that the better quality of EUV images by AIA-EUVI allows a more accurate detection of coronal holes in middle-low latitudes that are often under-represented or even missed by EIT. As a result, during 2010 - 2013, AIA-EUVI measures larger total hole area to be between 5-10\% of the total surface area, with dominant contribution from middle-low latitudes; it also measures significantly enhanced total open flux in the range of 2-4$\times 10^{22}$ Mx, which is about twice the flux measured by EIT, and matches with the PFSS calculated open flux. Nevertheless, there is a large discrepancy between AIA-EUVI measurements and PFSS results in terms of the latitude-dependence of coronal holes and open flux. PFSS appears to over-estimate the areas of coronal holes in the middle-latitudes that are extended from the polar holes, yet does not recover adequately holes in the lower latitudes as seen by AIA-EUVI.

To summarize, this comparative study of three different measurements (EIT, AIA-EUVI, and PFSS) of persistent coronal holes and open magnetic flux suggests that coronal holes in low-latitudes are important to contribute to total open flux. On the other hand, these holes are not well measured with either the chromosphere He I 10830 line in the past or EIT EUV images shown in the present study. Neither does the static PFSS model adequately produce lower latitude holes observed by AIA-EUVI, even though the measured and computed total open flux happen to match.

It has been discussed in the past that there is ambiguity in determining weaker or smaller coronal holes in the lower latitudes with the He I 10830 observations \citep{1982SoPh...79..203L}, and that coronal (SXR and EUV) and chromosphere (He I 10830) observations of these low-latitude holes often do not agree \citep{1983SoPh...87...47K,2003SoPh..212..165S, 2005SoPh..226....3M}. This discrepancy may result from an expansion of coronal holes with height. This study demonstrates that the AIA-EUVI EUV imagers are capable of uncovering many low-latitude holes compared with the EIT instrument. It should be noted that, given the analysis cadence of 24 hours for EIT and 12 hours for AIA/EUVI, most of these holes are long-lived persistent holes, as evidenced by the fact that the hole boundaries found in AIA images are very similar to those defined in STEREO-EUVI images.

From the point of view of models, \citet{2008ApJ...674.1158L} performed an MHD simulation of the steady-state magnetic field, and argued that PFSS reconstruction may have included active region contributions, and over-estimated the total steady-state open flux. Our observations comparing the latitude-dependence of open flux suggests that, whereas EIT measurements underestimate the low-latitude contribution, the PFSS may over-estimate this contribution by reproducing overall larger open areas than observed with AIA-EUVI. It was also put forward that the reconstructed total IMF, which is usually very large and comparable with the PFSS calculated flux, is due to the contribution of CME related open fluxes \citep{2007ApJ...667L..97R,2012SSRv..172..169A}. These events tend to be short-lived and occur near active regions in low-latitudes. To appropriately capture their contribution, it will be important to analyze the AIA-EUVI full surface observations at higher cadence. Even though the full-surface EUV observations are still limited by the one-side view of the magnetic field, tracking coronal hole evolution and detecting short-lived holes on the backside will provide indirect constraint to models that are used to calculate the open flux. For example, as suggested by \citet{2012LRSP....9....6M} in addition to other techniques to evolve the photospheric magnetic field \citep{2003SoPh..212..165S}, the boundary condition may be further controlled, or model assumptions varied, in advanced MHD models to reproduce observed coronal hole properties on the full surface.

\section{Acknowledgments}
We would like to thank the referee for constructive comments and suggestions. Particular thanks to Dana Longcope and Richard Canfield for discussion of this project. This work was supported by the NASA Living With a Star Program and NSF grant ATM-0748428. We acknowledge data use from the SDO, STEREO, and SoHO missions.

\end{document}